\documentclass[3p,twocolumn]{elsarticle}

\usepackage{epstopdf}
\usepackage[caption=false]{subfig}

\usepackage{amsmath,amsthm}

\usepackage{xcolor}
\usepackage{multirow}
\usepackage{kantlipsum}
\allowdisplaybreaks
\makeatletter
\newcommand{\mathleft}{\@fleqntrue\@mathmargin0pt}
\newcommand{\mathcenter}{\@fleqnfalse}
\makeatother

\bibpunct[, ]{[}{]}{,}{n}{,}{,}
\makeatletter
\def\NAT@def@citea{\def\@citea{\NAT@separator}}
\makeatother

\theoremstyle{plain}

\theoremstyle{definition}

\theoremstyle{remark}

\DeclareMathOperator{\erf}{erf}

\begin{document}
	\begin{frontmatter}

		\title{Removing non-smoothness in solving Black-Scholes equation using a 
			perturbation method}
		
		\author[laba]{Endah R.M.\ Putri\corref{cor1}}
		\ead{endahrmp@matematika.its.ac.id}
		\author[labb]{Lutfi Mardianto}
		\author[laba]{Amirul Hakam}
		\author[laba]{Chairul Imron}
		\author[labc,labd]{Hadi Susanto}
		
		\cortext[cor1]{Corresponding author.}
		\address[laba]{Department of Mathematics, Faculty of Mathematics, Computing, and Data Sciences, Institut Teknologi Sepuluh Nopember, Jl.\ Raya ITS, Sukolilo, Surabaya, 60111, Indonesia}
		\address[labb]{Department of Mathematics, Institut Teknologi Sumatera, Jl.\ Terusan Ryacudu, Way Hui, Jati Agung, Lampung Selatan, 35365 Indonesia}
		\address[labc]{Department of Mathematics, Khalifa University, Abu Dhabi Campus, PO.\ Box 127788, Abu Dhabi, United Arab Emirates}
		\address[labd]{Department of Mathematical Sciences, University of Essex, Wivenhoe Park, Colchester, CO4 3SQ, United Kingdom}
		
		
		\begin{abstract}
			{Black-Scholes} equation as one of the most celebrated mathematical models has an explicit analytical solution known as the Black-Scholes formula. Later variations of the equation, such as fractional or nonlinear Black-Scholes equations, do not have a closed form expression for the corresponding formula. In that case, one will need asymptotic expansions, including homotopy perturbation method, to give an approximate analytical solution. However, the solution is non-smooth at a special point. We modify the method by {first} performing variable transformations that push the point to infinity. As a test bed, we apply the method to the solvable Black-Scholes equation, where excellent agreement with the exact solution is obtained. We also extend our study to multi-asset basket and quanto options by reducing the cases to single-asset ones. Additionally we provide a novel analytical solution of the single-asset quanto option that is simple and different from the existing expression.

		\end{abstract}
		
		
	\end{frontmatter}
	
	\section{Introduction}
	\label{sec1}
	
	A central problem in financial derivative products for investment is their pricing or hedging. One of the most common products are options. An option is a financial contract which gives the holder a right, not obligation, to buy/sell underlying assets for certain price at maturity date.  It is of importance due to the use of option that is thriving in financial markets. Black-Scholes or Black-Scholes-Merton equation \cite{Black1973637,{Merton1973141}}, for {a European-style option valuation}, has been appreciated as one of the most celebrated mathematical models for its simplicity in giving a theoretical estimate of the option price and showing that it has a unique price regardless of the risk of the security.
	
	An exact solution of the equation, known as the Black-Scholes formula, has been derived in \cite{Black1973637,Merton1973141}. It was obtained analytically by solving the model as a diffusion equation, i.e., a parabolic partial differential equation. The formula can also be derived using, e.g., a Mellin transform \cite{pani04,joda05,agli13} or Green's function \cite{wilm95,hull06}.
	
	Approximate formulae have also been provided in terms of power series expansions using, {for example, the} Adomian decomposition method \cite{boh09,boh14}, homotopy perturbation method \cite{Gulkac20101349,Trachoo20171}, and a transformation method \cite{edek15}. Even though exact solutions of the problem have been obtained analytically, development of such approximations is necessary especially when one considers more complicated option pricing problems that do not admit solutions in simple closed forms. Nevertheless, there is an immediate shortcoming that the series approximation gives a non-smooth analytical solution at a single point, i.e., when the stock price is the same as the strike or exercise price. The problem arises because the property of the pay-off function, that is originally non-smooth, is carried over to the next orders of approximation. To overcome the limitation, some works consider differentiable, but rather cooked-up, pay-off functions, see for example \cite{avil13,gonz16,waki06}. However, such an approach may have less financial relevance from the application point of view.
	
	In this paper, we consider the Black-Scholes equation with the standard (non-smooth) pay-off function. {Firstly, we study the Black-Scholes equation for single-asset options.} {We then extend our study to more complicated cases, where basket and quanto options are discussed.} Herein, we limit ourselves to European put options for the sake of simplicity as the case for call options can be obtained easily by a put-call parity relation. 
	By applying a variable transformation that pushes the strike price to infinity, we show that the homotopy perturbation method will produce a smooth approximate analytical solution. Our result therefore improves that of \cite{Gulkac20101349} for the single-asset option. The same transformation is also applied to the basket and quanto options, where the pay-off for the former is a geometric mean of $n$-underlying assets, which is also non-smooth. The quanto option has a pay-off whose underlying asset is converted into another underlying asset at maturity. 
	
	In this paper, we also propose a new transformation to reduce the multi-asset quanto option into a single-asset one that allows us to obtain a simple analytical solution based on the Black-Scholes formula. To the best of our knowledge, the transformation is novel and the solution has not been reported before, i.e., a corresponding solution with a rather complicated expression in the form of multiple improper integrals was provided in \cite{jiang2005}. 
	
	As analytical solution of the equations considered herein is available, the reader may wonder as to what extent the homotopy perturbation method is still needed. We apply it to the solvable models as a test bed to demonstrate its applicability. In a following up paper, we will show that the method preceded by the proposed transformation also yields good approximations to the solution of Black-Scholes-type equations that have no known explicit expression. 

	The paper is presented using the following outlines. In Section \ref{sec2} we discuss the single-asset Black-Scholes model as the governing equation. In the same section, we also introduce the homotopy perturbation method and apply it to solve the main model following \cite{Gulkac20101349}. Additionally we point out several errors existing in \cite{Gulkac20101349}. In Section \ref{sec3} we discuss the variable transformation and perform the homotopy method. An extension of the study to multi-asset options are discussed in Section \ref{sec4} which consists of two parts: basket and quanto options. We compare the result of \cite{Gulkac20101349}, ours and the exact solution for a single asset in Section \ref{sec5} for the standard European options. Subsequently, the discussion about our results and analytical solutions of basket and quanto options, are also presented. Finally we conclude our work in Section \ref{sec6}.

	\section{Black-Scholes equation, homotopy, and the problem of non-smoothness}
	\label{sec2}
	
	The Black-Scholes differential equation for a single-asset European put option can be written as
	\begin{align}\label{eq:bssa}
	\frac{\partial P}{\partial t}+\frac{1}{2}\sigma^2S^2\frac{\partial^2 P}{\partial S^2}+rS\frac{\partial P}{\partial S}-rP=0,
	\end{align}
	where $P(S,t)$ is the value of a put option that depends on an underlying asset $S$ and time $t$.  Parameters in the model are volatility $\sigma$, strike price $E$, maturity date $T$, and a risk-free interest rate $r$ . We think of stock {as} the underlying asset in this paper. As the pay-off or final condition, we consider
	\begin{align*}
	P\left(S,T\right)=\max\left(K-S,\,0\right)
	\end{align*}
	and boundary conditions
	\begin{align*}
	P\left(0,t\right)=Ke^{-r(T-t)},\quad \lim_{S\rightarrow\infty}P(S,t)=0.
	\end{align*}
	
	
	By introducing the following dimensionless variables,
	\begin{align}
	S=Ke^x,\quad t=T-\frac{\tau}{\frac12 \sigma^2},\quad P=Kv\left(x,\tau\right),
	\label{tr1}
	\end{align}
	Eq.\ (\ref{eq:bssa}) can be transformed into the dispersion equation 
	\begin{align}
	\label{eq:bss}
	\frac{\partial v}{\partial\tau}-\frac{\partial^2 v}{\partial x^2}-\left(k-1\right)\frac{\partial v}{\partial x}+kv=0
	\end{align}
	with $k=2r/\sigma^2$. Equation (\ref{eq:bss}) only contains the dimensionless parameter $k$, representing the ratio between the interest rate and volatility of the stock return, and the dimensionless parameter time to expiry $\frac{1}{2}\sigma^2T$. Due to the variable transformation, the final condition becomes an initial one
	\begin{eqnarray}\label{eq:initial_condition}
	v_0(x,0)&=& \max \left(1-e^x,\,0\right).
	\end{eqnarray}
	The solution of the initial value problem \eqref{eq:bssa}, i.e., the Black-Scholes formula, can be written as
	\begin{equation}\label{exact}
	P\left(S,t\right)=Ee^{-r\left(T-t\right)}N\left(-d_2\right)-SN\left(-d_1\right),
	\end{equation}
	where $N(v)$ is a cumulative distribution function of a normal random variable
	\begin{equation}
	N(\upsilon)=\dfrac{1}{\sqrt{2\pi}}\int_{-\infty}^{\upsilon}e^{-t^2/2}dt,\label{Nv}
	\end{equation}
	with
	\begin{align*}
	d_1&=\frac{\log\left(S/E\right)+\left(r+\frac{1}{2}\sigma^2\right)\left(T-t\right)}{\sigma\sqrt{T-t}},\\
	d_2&=d_1 -\sigma\sqrt{T-t}.
	\end{align*}
	
	
	Gulka\c{c} \cite{Gulkac20101349} employed a homotopy perturbation method to solve the differential equation (\ref{eq:bss}) and (\ref{eq:initial_condition}). By adopting a homotopy technique, the method introduces a parameter in the system that initially is assumed to be small but later on is taken to be unity \cite{He200373,He1999257,odibat2008modified}. It is generally convergent, but one should be careful especially when the equation in consideration is nonlinear as convergence is not necessarily guaranteed as shown in \cite{Meijer2012323}.
	
	By using the method, we construct a homotopy equation \cite{Gulkac20101349}
	\begin{align}
	\label{eq:bstss}
	\frac{\partial v}{\partial \tau}=p\left(\frac{\partial^2v}{\partial x^2}+(k-1)\frac{\partial v}{\partial x}-kv\right).
	\end{align}
	Note that Eq.\ (\ref{eq:bss}) is obtained from \eqref{eq:bstss} by taking $p=1$. We seek for the solution of Eq.\ (\ref{eq:bstss}) in the form of the power series \cite{Gulkac20101349}
	\begin{align}
	\label{eq:sols}
	v=v_0+pv_1+p^2v_2+\dots,
	\end{align}
	where $v_0$ is given by \eqref{eq:initial_condition}. Substituting it into \eqref{eq:bss} and collecting terms with the same power will yield at $\mathcal{O}(p^n)$, $n=1,2,3,\dots,$
	\begin{align}
	\begin{split}
	\frac{\partial v_n}{\partial\tau}=\frac{\partial^2 v_{n-1}}{\partial x^2}+\left(k-1\right)\frac{\partial v_{n-1}}{\partial x}-kv_{n-1}. 
	\end{split} \label{tamb}\end{align}
	
	The problem with the homotopy perturbation method appears here. Because $v_0$ {in Eq.\ (\ref{eq:initial_condition})} is not smooth at $x=0$, that corresponds to the strike price $S=E$, the function is not differentiable at that point. To overcome {the non-differentiability}, the computation region is normally split into two parts, i.e., $e^x<1$ and $e^x\geq1$. Solving \eqref{tamb} in the respective region yields 
	\begin{align}
	v_{n}(x,\tau)&=
	\begin{cases}
	\dfrac{\left( -k\tau\right) ^n}{n!}, &e ^x<1,\\
	0,    & e^x\geq 1.
	\end{cases}
	\end{align}
	By taking $p\rightarrow 1$ in \eqref{eq:sols} and recognising that $\sum_{n=1}^{\infty}{\left( -k\tau\right) ^n}/{n!}=e^{-k\tau}-1$, the {put} option value is finally obtained as
	\begin{align}
	v(x,\tau)=\begin{cases}
	\begin{array}{cc}
	-e^x+e^{-k\tau}, & e^x<1{,} \\
	0, &  e^x\geq1.
	\end{array}
	\end{cases}
	\label{eq:solbss}
	\end{align}
	The solution Eq.\ (\ref{eq:solbss}) has the non-smoothness problem at $x=0$ that we explained in Section \ref{sec1}. In fact, the solution is not even continuous at that point for $\tau>0$. Therefore, when we plot the solution obtained from the homotopy perturbation method, we approximate it with $v(x,\tau)=\max(-e^x+e^{-k\tau},0)$.
	
	It is important to give a remark that there is a flaw in the calculations and result of \cite{Gulkac20101349}. The final solution (option value) given as Eq.\ (28) therein is not correct because it does not satisfy the initial condition (pay-off function). This is caused by the mistake in evaluating the differential equation for $v_1$ (see Eq.\ (20) therein), where $\partial v_0/\partial\tau$ should have been taken to be zero that leads to the wrong constant of integration in the expression of $v_1$ (see Eq.\ (23) therein).
	
	\section{Homotopy perturbation method with a variable transformation}
	\label{sec3}
	
	In this section, we present a way to remove the non-smoothness in the solution obtained using the homotopy perturbation method. To do so, we begin with applying the following variable transformations \cite{ke2018appropriate}
	\begin{equation}
	\label{TransKe}
	z=\dfrac{x}{\sqrt{\tau}}, \qquad w=\sqrt{\tau}, \qquad u=\dfrac{v}{\sqrt{\tau}},
	\end{equation}
	such that Eq.\ (\ref{eq:bss}) can be rewritten as
	\begin{align}
	\label{eq:bss1}
	\dfrac{\partial (wu)}{\partial w}=2\dfrac{\partial^2 u}{\partial z^2}+z\dfrac{\partial u}{\partial z}+2\left(k-1\right)w\dfrac{\partial u}{\partial z}-2kw^2u.
	\end{align}
	Note that due to the transformation, point $x=0$ where non-smoothness is located at $\tau=0$ is now shifted to infinity, i.e., $z=\lim\limits_{\tau\rightarrow0}\dfrac{x}{\sqrt{\tau}}=\pm\infty$. Accordingly, the initial condition (\ref{eq:initial_condition}) now becomes
	\mathleft
	\begin{align}
	\label{eq:Taylor}
	&\lim_{w\rightarrow0}u(z,w)\\
	\nonumber
	&=
	\begin{cases}
	\dfrac{1-e^{zw}}{w}=-z-\dfrac{z^2}{2!}w-\dfrac{z^3}{3!}w^2-\dots,\hspace{-7pt}&z\rightarrow -\infty \\
	0,&z\rightarrow \infty.
	\end{cases}
	\end{align}
	
	In order to solve the partial differential equation (\ref{eq:bss1}) with the 'initial condition' \eqref{eq:Taylor} using the homotopy perturbation method, we construct a homotopy equation
	\begin{equation}\label{eq:Homp}
	\dfrac{\partial\left(wu\right)}{\partial w}=2\dfrac{\partial^2 u}{\partial z^2}+z\dfrac{\partial u}{\partial z}+2p(k-1)w\dfrac{\partial u}{\partial z}-2p^2kz^2u.\\
	\end{equation}
	Again we assume that Eq.\ (\ref{eq:Homp}) has a power series solution
	\mathcenter
	\begin{equation}\label{eq:limPower}
	u=u_0+pu_1+p^2u_2+\dots.
	\end{equation}
	Again, substituting \eqref{eq:limPower} into \eqref{eq:Homp} and collecting terms with the same power result in
	\begin{align}
	\begin{split}
	p^0~:&~2\dfrac{\partial^2 u_0}{\partial z^2}+z\dfrac{\partial u_0}{\partial z}-\dfrac{\partial (wu_0)}{\partial w}=0,\\
	p^1~:&~2\dfrac{\partial^2 u_1}{\partial z^2}+z\dfrac{\partial u_1}{\partial z}-\dfrac{\partial (wu_1)}{\partial w}+2(k-1)w\dfrac{\partial u_0}{\partial z}=0,\\
	&\vdots\\
	p^n~:&~2\dfrac{\partial^2 u_n}{\partial z^2}+z\dfrac{\partial u_n}{\partial z}-\dfrac{\partial (wu_n)}{\partial w}+2(k-1)w\dfrac{\partial u_{n-1}}{\partial z}\\
	&-2kw^2u_{n-2}=0,
	\end{split}\label{eexp}
	\end{align}
	with $n=2,3,\dots.$
	
	A little inspection on the initial condition (\ref{eq:Taylor}) suggests us that we should look for solutions in the form of
	\begin{equation}
	u_i(z,w)=f_i(z)w^i,\quad i=0,1,2,\dots,
	\label{ser1}
	\end{equation}
	with boundary conditions
	\begin{eqnarray}\label{eq:BCon}
	f_i(z)=\begin{cases}
	-\dfrac{z^{i+1}}{(i+1)!}, & z\rightarrow-\infty,\\
	0, & z\rightarrow\infty.
	\end{cases}
	\end{eqnarray}
	
	Solving the resulting differential equations from \eqref{eexp} and then using the boundary conditions \eqref{eq:BCon} yield
	\begingroup
	\allowdisplaybreaks
	\begin{subequations}
		\label{soluNoSingularity}
		\begin{align}
		u_0&(z,w)=\dfrac{e^{-\dfrac{z^2}{4}}}{\sqrt{\pi}}+\dfrac{z}{2}\left(\erf\left(\dfrac{z}{2}\right)-1\right),\\
		u_1&(z,w)=\dfrac{w}{4}\left[\dfrac{2ze^{-\dfrac{z^{2}}{4}}}{\sqrt{\pi}}+\left(z^2+2k\right)\left(\erf\left(\dfrac{z}{2}\right)-1\right)\right],\\
		\nonumber
		u_2&(z,w)=\dfrac{w^{2}}{12}\left[\dfrac{e^{-\dfrac{z^{2}}{4}}}{\sqrt{\pi}}\left( 2z^{2}+3k^{2}-6k-1\right)\right.\\
		&\left.\left.+z^{3}\left(\erf\left(\dfrac{z}{2}\right)-1\right)\right]\right.,\\
		\nonumber
		u_3&(z,w)=\dfrac{w^3}{48}\left[\dfrac{2ze^{-\dfrac{z^{2}}{4}}}{\sqrt{\pi}}\left(z^2-k^3+3k^2-3k-1\right)\right.\\
		&\left.+\left(z^4-12k^2\right)\left(\erf\left(\dfrac{z}{2}\right)-1\right)\right],\\
		\nonumber
		u_4&(z,w)=\dfrac{w^4}{960}\left[\dfrac{e^{-\dfrac{z^{2}}{4}}}{\sqrt{\pi}}\left(8z^4+(5k^4-20k^3+30k^2\right.\right.\\\nonumber
		&-20k-11)z^2-10k^4-120k^3+180k^2\\
		&\left.\left.+40k+6\right)+4z^5\left(\erf\left(\dfrac{z}{2}\right)-1\right)\right],\\\nonumber
		u_5&(z,w)=\dfrac{w^5}{5760}\left[\dfrac{e^{-\dfrac{z^{2}}{4}}}{\sqrt{\pi}}\left(8z^5-(3k^5-15k^4+30k^3\right.\right.\\\nonumber
		&-30k^2+15k+13)z^3\\\nonumber
		&\left.+(18k^5+90k^4-300k^3+180k^2+90k+18)z\right)\\
		&\left.+\left(4z^6+480k^3\right)\left(\erf\left(\dfrac{z}{2}\right)-1\right)\right].
		\end{align}
	\end{subequations}
	
	\endgroup
	We only present the first five terms of the solution as they will be sufficient to show the significant improvement in our approximate solution using the homotopy perturbation method with the additional transformation presented above. 

	\section{Extensions on multi-asset options}\label{sec4}
	
	Options developed by two or more underlying assets are called multi-asset options and the price satisfies multidimensional parabolic differential equations. The different types of options are characterized by their pay-off structures. Basket options have their pay-off as the geometric mean of the underlying assets, while the pay-off of quanto options converts one underlying asset into another one at maturity. 
	
	The Black-Scholes differential equation for multi-asset options 
	can be written as
	\begin{align}\label{maput2}
	\nonumber
	\frac{\partial P}{\partial t}&+\frac{1}{2}\sum_{i,j=1}^{n}{a_{ij}S_iS_j\frac{\partial^2 P}{\partial S_i S_j}}\\
	&+\sum_{i=1}^{n}{(r-q_i)S_i\frac{\partial P}{\partial S_i}}-rP=0,
	\end{align}
	where $q_i$ is a dividend rate of the underlying asset $S_i$,
	\begin{align*}
	a_{ij}&=\sum_{k=1}^{m}\sigma_{ik}\sigma_{jk}, \,\, (i,j=1,\dots,n),
	\end{align*}
	and $\sigma_{ij}$ is the volatility of return of asset $(i,j)$.
	
	\subsection{Basket options}\label{sec4a}
	
	The basket option governing equation refers to Eq.\ (\ref{maput2}) with its pay-off function given by 
	\begin{align}
	P({S_1,S_2,\cdots,S_n},T)=\max \left(K-\prod_{i=1}^{n} S_{i}^{\alpha_{i}}\right).\label{pfbasket0}
	\end{align}
	Introducing similar transformations to those in Eq.\ (\ref{tr1}) with some adjustments for the multi-underlying assets as used in, e.g., \citep{jiang2005}: 
	\begin{alignat*}{2}
	S_i&=Ke^{x_i};\qquad&&t=T-\frac{\tau}{\frac12 \hat{\sigma}^2};\\
	P&=Kv(x,\tau);\qquad&&\xi=\sum_{i}^{n}{\alpha_{i}x_{i}}
	\end{alignat*}
	where $\hat{\sigma}^2=\sum_{i,j=1}^{n}{a_{ij}\alpha_{i}\alpha_{j}}$ and $\sum_{i}^{n}{\alpha_{i}}=1$, Eq.\ (\ref{maput2}) can then be simplified into the following equation similar to the single-asset option in Sec.\ \ref{sec2},
	\begin{align}
	\label{maput3}
	\frac{\partial v}{\partial \tau}= \frac{1}{2}{\hat{\sigma}}^2 \frac{\partial^2 v}{\partial \xi^2}+\left( r-\hat{q}-\frac{1}{2}\hat{\sigma}^2 \right) \frac{\partial v}{\partial \xi}-rv,
	\end{align}
	where 
	\begin{align}
	\hat{q}&=\sum_{i=1}^{n}{\alpha_{i}\left(q_i+\frac{a_{ii}}{2}\right)}-\frac{\hat{\sigma}^2}{2}.
	\end{align}
	and the pay-off function (\ref{pfbasket0}) becomes an initial condition,  
	\begin{align}
	v\left(\xi,0\right)&=\text{max}\left(1-e^{\xi},0\right).\label{pfbasket}
	\end{align}
	Note that the exact solution of basket put options for two assets are given by \cite{jiang2005}
	\newpage
	\begin{align}
	\label{sol2}
	\nonumber
	P(S_1,S_2,t)&=Ee^{-r(T-t)}N(-\hat{d_2})\\
	&-e^{-\hat{q}(T-t)}S_1^{\alpha_1}S_2^{\alpha_2}N(\hat{-d_1}),
	\end{align}
	with
	\mathleft
	\begin{align*}
	\hat{d_1}&=\dfrac{\ln{\dfrac{S_1^{\alpha_1}S_2^{\alpha_2}}{E}}+\left[r-\hat{q}+\frac{\hat{\sigma}^2}{2}\right](T-t)}{\hat{\sigma}\sqrt{(T-t)}}\\
	\hat{d_2}&=\hat{d_1}-{\hat{\sigma}\sqrt{(T-t)}},\\
	\hat{q}&=\sum_{i=1}^{n}{\alpha_{i}\left(q_i+\frac{a_{ii}}{2}\right)}-\frac{\hat{\sigma}^2}{2},\\
	\hat{\sigma}^2&=\sum_{i,j=1}^{n}{a_{ij}\alpha_{i}\alpha_{j}},\\
	\sum_{i}^{n}{\alpha_{i}}&=1.
	\end{align*}
	
	In solving the multi-asset basket option using the homotopy perturbation method, note that the function is also not smooth. The same transformation (\ref{TransKe}) is then applied here to 'push' the point of non-smoothness to infinity, i.e.,
	\mathcenter
	\begin{equation}
	z=\dfrac{\xi}{\sqrt{\tau}}, \qquad w=\sqrt{\tau}, \qquad u=\dfrac{v}{\sqrt{\tau}}.
	\end{equation}
	We therefore construct the following homotopy equation
	\begin{align}
	\label{bc}
	\nonumber
	\frac{\partial \left(uw\right)}{\partial w}&=\hat{\sigma}^2\frac{\partial^2 u}{\partial z^2}+z\frac{\partial u}{\partial z} +2p\left(r-\hat{q}-\frac{1}{2}\hat{\sigma}^2\right)\frac{\partial u}{\partial z}w\\
	&-2p^2ruw^2
	\end{align}
	The initial condition (\ref{pfbasket}) becomes 
	\begin{align}
	&\lim\limits_{w \rightarrow 0}u(z,w)\\
	\nonumber
	&=
	\begin{cases}
	\dfrac{1-e^{zw}}{w}=-z-\dfrac{z^2}{2!}w-\dfrac{z^3}{3!}w^2-\dots,\hspace{-7pt}&z\rightarrow -\infty \\
	0,&z\rightarrow \infty.
	\end{cases}
	\end{align}
	
	The next step is to solve the partial differential equation (\ref{bc}) using the homotopy perturbation method as before. Writing 	\begin{equation}\label{tam1}
	u=u_0+pu_1+p^2u_2+\dots.
	\end{equation}
	and performing the same procedures, we obtain 
	\begin{subequations}
		\begin{align}
		u_0&(z,w)=\dfrac{e^{-\dfrac{z^2}{4}}}{\sqrt{\pi}}+\dfrac{1}{2}z\left(\erf\left(\dfrac{z}{2}\right)-1\right),\\
		\nonumber
		u_1&(z,w)
		=\dfrac{w}{4}\left[\dfrac{2e^{-\dfrac{z^2}{4}}}{\sqrt{\pi}}z+\frac{1}{4\hat{\sigma}^2}\left(\hat{\sigma}^2z^2-4(\hat{q}-r)\right)\right.\\
		&\left.\left(\erf\left(\dfrac{z}{2}\right)-1\right)\right],\\
		\nonumber
		u_2&(z,w)=\dfrac{w^2}{12}\left[\dfrac{e^{-\dfrac{z^2}{4}}}{\hat{\sigma}^4\sqrt{\pi}}\left(\hat{\sigma}^4(2z^2-1)\right.\right.\\
		\nonumber
		&\left.-12\hat{q}(\hat{\sigma}^2+2r)-12\hat{\sigma}^2r+12\hat{q}^2+12r^2\right)\\
		&{\left.+\dfrac{z}{\hat{\sigma}^2}(\hat{\sigma}^2z^2-12\hat{q})\left(\erf\left(\dfrac{z}{2}\right)-1\right)\right],}
		\\
		\nonumber
		u_3&(z,w)=\dfrac{w^3}{48}\left[\dfrac{2e^{-\dfrac{z^2}{4}}}{\hat{\sigma}^6\sqrt{\pi}}z\left(\hat{\sigma}^6z^2-2\hat{q}(9\hat{\sigma}^2\right.\right.\\\nonumber
		&\left.-6\hat{\sigma}^2\hat{q}-4\hat{q}^2\right)-6r(\hat{\sigma}^2+2\hat{q})+12r^2(\hat{\sigma}^2+2\hat{q})\\
		\nonumber
		&\left.-\hat{\sigma}^6-8r^3\right)+\frac{1}{\hat{\sigma}^4}\left(\hat{\sigma}^4z^4-24\hat{\sigma}^2\hat{q}z^2+48\hat{q}-48r^2\right)\\
		&\left.\left(\erf\left(\dfrac{z}{2}\right)-1\right)\right],\\
		\nonumber
		u_4&(z,w)
		\nonumber
		=\dfrac{w^4}{960}\left[\dfrac{e^{-\dfrac{z^2}{4}}}{\hat{\sigma}^8\sqrt{\pi}}\left(8\hat{\sigma}^8z^4-\left(11\hat{\sigma}^8\right.\right.\right.\\
		\nonumber
		&+40\hat{\sigma}^6(7\hat{q}+r)-120\hat{\sigma}^4(\hat{q}-r)^2-160\hat{\sigma}^2(\hat{q}-r)^3\\
		\nonumber
		&\left.-80(\hat{q}-r)^4\right)z^2+6\hat{\sigma}^8+80\hat{\sigma}^6(\hat{q}+r)\\
		\nonumber
		&+240\hat{\sigma}^4(3\hat{q}^2+2\hat{q}r+3r^2)-960\hat{\sigma}^2(\hat{q}-r)^2(\hat{q}+r)\\
		\nonumber
		&-160\left(\hat{q}-r)^4\right)\\
		\nonumber
		&\left.+\frac{4}{\hat{\sigma}^4}z\left(\hat{\sigma}^4z^4-40\hat{\sigma}^2\hat{q}z^2+240\hat{q}^2\right)\left(\erf\left(\dfrac{z}{2}\right)-1\right)\right],\\
		\nonumber
		u_5&(z,w)=\dfrac{w^5}{5760}\left[\dfrac{e^{-\dfrac{z^2}{4}}}{\hat{\sigma}^{10}\sqrt{\pi}}\left(8\hat{\sigma}^{10}z^5-\left(13\hat{\sigma}^{10}\right.\right.\right.\\\nonumber
		&+30\hat{\sigma}^8(15\hat{q}+r)-120\hat{\sigma}^6(\hat{q}-r)^2-240\hat{\sigma}^4(\hat{q}-r)^3\\
		\nonumber
		&\left.-240\hat{\sigma}^2(\hat{q}-r)^4-96(\hat{q}-r)^5\right)z^3+\left(18\hat{\sigma}^{10}\right.\\
		\nonumber
		&{+60\hat{\sigma}^8(5\hat{q}+3r)+720\hat{\sigma}^6(5\hat{q}^2+2\hat{q}r+r^2)}\\
		\nonumber
		&-480\hat{\sigma}^4(\hat{q}-r)^2(7\hat{q}+5r)\\\nonumber
		&\left.\left.-480\hat{\sigma}^2(\hat{q}-r)^3(5\hat{q}+3r)-576(\hat{q}-r)^5\right)z\right)\\\nonumber
		&+\dfrac{4}{\hat{\sigma}^6}\left(\hat{\sigma}^6z^6-60\hat{\sigma}^4\hat{q}z^4+720\hat{\sigma}^2\hat{q}^2z^2\right.\\
		&\left.\left.-960\hat{q}^3+960r^3\right)\left(\erf\left(\dfrac{z}{2}\right)-1\right)\right].
		\end{align}
		\label{tam4}
	\end{subequations}
	We do not continue the computation further. 
	\subsection{Quanto options}\label{sec4b}
	
	A quanto option is a short term of a quantity-adjusting option in which the underlying assets are valued in a different currency from the currency that the investors settle. In this case, investors invest in options with foreign underlying assets but keep the payout in their home currency. A greater liquidity obtained by removing currency risks is the benefit of this option. The governing equation of quanto options also refers to Eq.\ (\ref{maput2}) with several different features. 
	
	There are two types of multi-underlying assets in this option: the underlying asset in a foreign currency in which the option is issued that is denoted by $S_1$ and the exchange rate ratio between home and the foreign currency denoted as $S_2$. 
	If the investors have a portfolio in a foreign currency, then it will consist of a long position on one option and a short one on a number of the underlying assets $S_1$, adjusted by the home currency $S_2$ (i.e., we obtain $S_2S_1$) and the foreign exchange rate ratio $S_2$. By treating the first short position as a new underlying asset, denoted as $\hat{S}_1$, then we can consider Eq.\ (\ref{maput2}) as a multi-asset governing equation for quanto options. After mathematical modifications, {after reverting $\hat{S}_1$ back to} its original form (see \cite{jiang2005} for the details), the partial differential equation for the options can be written as 
	\begin{align}
	\label{eq1}
	\nonumber
	\frac{\partial P}{\partial t}&+\frac{1}{2}\left(\sigma_1^2S_1^2\frac{\partial^2P}{\partial S_1^2}+2\rho\sigma_1\sigma_2S_1S_2\frac{\partial^2P}{\partial S_1\partial S_2}\right.\\
	\nonumber
	&\left.+\sigma_2^2S_2^2\frac{\partial^2P}{\partial S_2^2}\right)+\left(r_1-q-\rho\sigma_1\sigma_2\right)S_1\frac{\partial P}{\partial S_1}\\
	\nonumber
	&+\left(r_1-r_2\right)S_2\frac{\partial P}{\partial S_2}+\left(r_2-r_1\right)S_1\frac{\partial P}{\partial S_1}\\
	&-r_1P=0.
	\end{align}
	The payoff function 
	is defined as{
		\begin{align}
		P(S_1,S_2,T)=S_2(T)\max\left(E-S_1(T),0\right).
		\end{align}}
	To obtain the solution of the options, 
	we propose a new transformation to convert Eq.\ (\ref{eq1}) to a single-asset one, i.e., 
	\begin{align*}
	v=\frac{P}{S_2^2},\,
	x=\frac{S_1}{S_2}.
	\end{align*}
	Subsequently, the payoff function after the transformation can be written as 
	{
		\begin{align*}
		v(x,T)=\max\left(K-x,0\right),
		\end{align*}}
	{where} $K=E/S_2(T)$.
	
	Applying the variable transformation to Eq.\ (\ref{eq1}) yields a single-asset equation that is similar to Eq.\ (\ref{maput3}),
	\mathcenter
	\begin{align}\label{quantosingle}
	\frac{\partial v}{\partial t}+\frac{1}{2}\hat{\sigma}^2x^2\frac{\partial^2v}{\partial x^2}+\hat{q}x\frac{\partial v}{\partial x}-\hat{r}v=0,
	\end{align}
	where 
	\begin{align*}
	\hat{\sigma}^2&=\sigma_1^2-2\rho\sigma_1\sigma_2+\sigma_2^2,\\
	\hat{q}&=2r_2-r_1-q-\sigma_2^2,\\
	\hat{r}&=r_1-2r_2+\sigma_2^2.
	\end{align*}
	
	Next, we define the following dimensionless variables 
	\begin{align*}
	x=Ke^y,\quad v=Ku(y,\tau),\quad t=T-\frac{\tau}{\frac{1}{2}\hat{\sigma}^2},
	\end{align*}
	{and} apply them to Eq.\ (\ref{quantosingle}) to yield 
	\begin{align}
	\label{eq3}
	\frac{\partial u}{\partial\tau}=\frac{\partial^2u}{\partial y^2}+(k_1-1)\frac{\partial u}{\partial y}-k_2u=0,	
	\end{align}
	where $k_1=2\hat{q}/\hat{\sigma}^2$ and $k_2=2\hat{r}/\hat{\sigma}^2$.
	Accordingly, the payoff function as a final condition now becomes an initial condition
	{
		\begin{align}\label{payoffQuanto}
		u(y,0)=\max\left(1-e^y,0\right).
		\end{align}}
	In the following, we will provide the explicit solution of the problem and an approximate one using a homotopy perturbation method.
	
	\subsubsection{Exact solution for the quanto options}\label{sec4b1}
	
	To solve the "single-asset" quanto option  Eq.\ (\ref{eq3}), we use a common transformation discussed in the literature \cite{wilm95, jiang2005} that will simplify the differential equation, namely 
	\begin{align*}
	u=e^{\alpha\tau+\beta y}w(y,\tau),
	\end{align*}
	which upon substitution into Eq.\ (\ref{eq3}) and choosing $\beta=-\frac{k_1-1}{2}$ and $\alpha=-\frac{(k_1-1)^2}{4}-k_2$, will yield 
	\begin{align}
	\label{eq:heat}
	w_\tau-w_{yy}=0,
	\end{align}
	and the initial condition 
	\begin{align}
	w(y,0)&=e^{-\alpha\tau-\beta y}u(y,0)
	\nonumber\\
	&=\text{max}\left(e^{\left(\frac{k_1-1}{2}\right)y}-e^{\left(\frac{k_1+1}{2}\right)y},0\right).
	\label{tam2}
	\end{align}
	
	The solution of the Cauchy problem (\ref{eq:heat}) and \eqref{tam2} is 
	\begin{align*}
	w(y,\tau)=\int_{-\infty}^{\infty}H(y-s,\tau)w(s,0)\,ds,
	\end{align*}
	where $H(y-s,\tau)$ is the fundamental solution of the heat equation,
	\begin{align*}
	H(y-s,\tau)=\frac{1}{2\sqrt{\pi\tau}}e^{-\frac{\left(y-s\right)^2}{4\tau}}.
	\end{align*}
	
	By taking $\omega=\frac{s-y}{\sqrt{2\tau}}$ and thereby $d\omega=\frac{1}{\sqrt{2\tau}}ds$, the solution can be written as
	\begin{align}\label{eq:transQuanto}
	w(y,\tau)&=\frac{1}{\sqrt{2\pi}}\int_{-\infty}^{\infty}e^{-\frac{\omega^2}{2}}w(\omega\sqrt{2\tau}+y,0)d\omega\nonumber\\
	&=\frac{1}{\sqrt{2\pi}}\int_{-\infty}^{-\frac{y}{\sqrt{2\tau}}}e^{-\frac{\omega^2}{2}}\left(e^{\left(\frac{k_1-1}{2}\right)\left(\omega\sqrt{2\tau}+y\right)}\right.\nonumber\\
	&\left.-e^{\left(\frac{k_1+1}{2}\right)\left(\omega\sqrt{2\tau}+y\right)}\right)d\omega\nonumber\\
	&=I_1+I_2.
	\end{align}
	
	The first term $I_1$ defined as
	\begin{align}\label{I1}
	I_1=\frac{1}{\sqrt{2\pi}}\int_{-\infty}^{-\frac{y}{\sqrt{2\tau}}}e^{-\frac{\omega^2}{2}+\left(\frac{k_1-1}{2}\right)\left(\omega\sqrt{2\tau}+y\right)}d\omega
	\end{align}
	can be rearranged to 
	obtain
	\mathcenter
	\begin{align}\label{IntegralI1}
	I_1&=\frac{1}{\sqrt{2\pi}}e^{\frac{k_1-1}{2}y+\frac{(k_1-1)^2}{4}\tau}\int_{-\infty}^{\frac{y}{\sqrt{2\tau}}}e^{-\frac{1}{2}\left(\omega-\sqrt{\frac{\tau}{2}}(k_1-1)\right)^2}d\omega\nonumber\\
	&=e^{\frac{k_1-1}{2}y+\frac{(k_1-1)^2}{4}\tau}N(-d_1),
	\end{align}
	where 
	\begin{align*}
	d_1&=\frac{y}{\sqrt{2\tau}}+\sqrt{\frac{\tau}{2}}(k_1-1)\\
	&=\frac{\log(S_1/E)+\left(\hat{q}-\frac{\hat{\sigma}^2}{2}\right)(T-t)}{\hat{\sigma}\sqrt{T-t}}.
	\end{align*}
	
	The second integral $I_2$ can also be simplified by using the same procedure into 
	\begin{align*}
	I_2=-e^{\frac{k_1+1}{2}y+\frac{(k_1+1)^2}{4}\tau}N(-d_2),
	\end{align*}
	where
	\begin{align*}
	d_2&=\frac{y}{\sqrt{2\tau}}+\sqrt{\frac{\tau}{2}}(k_1+1)\\
	&=\frac{\log(S_1/E)+\left(\hat{q}+\frac{\hat{\sigma}^2}{2}\right)(T-t)}{\hat{\sigma}\sqrt{T-t}}.
	\end{align*}
	
	Reverting back all the transformed variables, we finally obtain 
	the analytic solution of the quanto put option as
	\begin{align}\label{eq:analyticquanto}
	P&=ES_2e^{-\hat{r}(T-t)}N(-d_1)-S_1S_2e^{(\hat{q}-\hat{r})(T-t)}N(-d_2),
	\end{align}
	which to our best knowledge has never been reported before. 
	
	\subsubsection{A homotopy perturbation method for quanto options}\label{4b2}
	
	In a similar fashion as in the previous sections, we will also derive an asymptotic solution of the quanto options using the homotopy method. 
	Again we apply the same variable transformations \eqref{TransKe}, that in here are given by
	\begin{align*}
	\xi=\frac{y}{\sqrt{\tau}},\quad z=\sqrt{\tau},\quad\varphi=\frac{u}{\sqrt{\tau}}.
	\end{align*}
	Equation (\ref{eq3}) now becomes
	\begin{align}
	\frac{\partial(z\varphi)}{\partial z}&=2\frac{\partial^2\varphi}{\partial\xi^2}+\xi\frac{\partial\varphi}{\partial\xi}+2(k_1-1)z\frac{\partial\varphi}{\partial\xi}-2k_2z^2\varphi.
	\end{align}
	By assuming that the solution $\phi$ can be written in a series form as $\phi=\phi_0+\phi_1+\phi_2+\dots$, we obtain
	\begin{subequations}
		\begin{align}
		\label{eq:hpmquanto1}
		\varphi_0&(\xi,z)=\dfrac{e^{-\dfrac{\xi^2}{4}}}{\sqrt{\pi}}+\dfrac{1}{2}\xi\left(\erf\left(\dfrac{\xi}{2}\right)-1\right)\\
		\label{eq:hpmquanto2}
		\varphi_1&(\xi,z)=\dfrac{z}{4}\left[\dfrac{2e^{-\dfrac{\xi^2}{4}}}{\sqrt{\pi}}\xi+\left(\xi^2+2k_1\right)\left(\erf\left(\dfrac{\xi}{2}\right)-1\right)\right]\\
		\label{eq:hpmquanto3}
		\nonumber
		\varphi_2&(\xi,z)=\dfrac{z^2}{12}\left[\dfrac{e^{-\dfrac{\xi^2}{4}}}{\sqrt{\pi}}\left(2\xi^2+3k_1^2+6k_1-12k_2-1\right)\right.\\
		&\left.+\xi\left(\xi^2+6k_1-6k_2\right)\left(\erf\left(\dfrac{\xi}{2}\right)-1\right)\right]\\
		\nonumber
		\varphi_3&(\xi,z)=\dfrac{z^3}{48}\left[\dfrac{2e^{-\dfrac{\xi^2}{4}}}{\sqrt{\pi}}\xi\left(\xi^2-k_1^3+3k_1^2+9k_1\right.\right.\\
		\nonumber
		&\left.\left.-12k_2-1\right)+\left(\xi^4+12(k_1-k_2)\xi^2\right.\right.\nonumber\\
		&\left.\left.+12k_1(k_1-2k_2)\right)\left(\erf\left(\dfrac{\xi}{2}\right)-1\right)\right]\\
		\label{eq:hpmquanto4}
		\nonumber
		\varphi_4&(\xi,z)=\dfrac{z^4}{960}\left[\dfrac{e^{-\dfrac{\xi^2}{4}}}{\sqrt{\pi}}\left(8\xi^4+(5k_1^4-20k_1^3+30k_1^2\right.\right.\\
		\nonumber
		&+140k_1-11)\xi^2-10k_1^4+120k_1^3+180k_1^2\nonumber\\
		&\left.-40k_1-240(k_1^2+2k_1)k_2+480k_2^2+80k_2+6\right)\nonumber\\
		&+4\xi\left(\xi^4+20(k_1-k_2)\xi^2+60(k_1-k_2)^2\right)\nonumber\\
		&\left.\left(\erf\left(\dfrac{\xi}{2}\right)-1\right)\right]\\
		\nonumber
		\label{eq:hpmquanto5}
		\varphi_5&(\xi,z)=\dfrac{z^5}{5760}\left[\dfrac{e^{-\dfrac{\xi^2}{4}}}{\sqrt{\pi}}\left(8\xi^5-(3k_1^5-15k_1^4+30k_1^3\right.\right.\\
		\nonumber
		&\left.-30k_1^2-225k_1+240k_2+13)\xi^3\right.\\
		\nonumber
		&+(18k_1^5-150k_1^4+420k_1^3+900k_1^2-150k_1\\
		\nonumber
		&+(240k_1^3-720k_1^2-2160k_1)k_2\\
		\nonumber
		&\left.+240k_2+1440k_2^2+18)\xi\right)\\
		\nonumber
		&+4\left(\xi^6+30(k_1-k_2)\xi^4+180(k_1-k_2)^2\xi^2\right.\\
		&\left.\left.+120k_1(k_1^2-3k_1k_2+3k_2^2)\right)\left(\erf\left(\dfrac{\xi}{2}\right)-1\right)\right].
		\end{align}
	\end{subequations}
	One can continue computing the next order solutions, which are left to the interested reader. 
	
	\section{Discussion}
	\label{sec5}
	
	\subsection{Single-asset European put options}\label{sec5a}
	
	In this section, we compare the analytical results obtained in Sec.\ \ref{sec3} with those in \cite{Gulkac20101349}. We call the results in \cite{Gulkac20101349} which contain the non-smoothness problem as HPM1 and our results Eqs.\ (\ref{soluNoSingularity}) and \eqref{eq:limPower} (with $p=1$) as HPM2, respectively. To show the accuracy of our results, we also compare them with the exact solution \eqref{exact}.
	\begin{figure}[tbhp!]
		\mbox{}\hfill
		\includegraphics[scale=0.4]{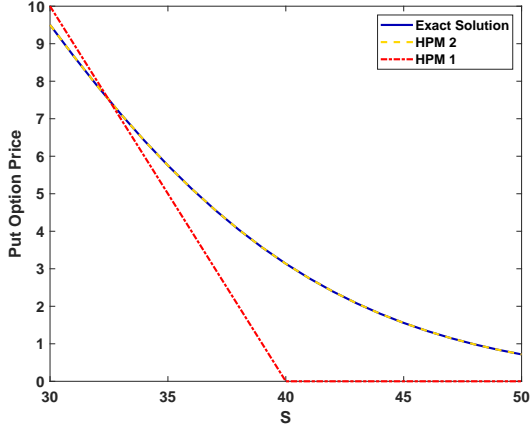}
		\hfill~
		\caption{Plot of our approximations (dash-dotted and dashed lines) and the exact solution \eqref{exact} (solid line). Dash-dotted line is HPM1, which is obtained from applying the inverse transformation of \eqref{tr1} to the solution \eqref{eq:solbss}. The dashed line is HPM2, which is obtained from \eqref{eq:limPower}, \eqref{ser1}, and \eqref{soluNoSingularity} with the transformation \eqref{TransKe} and \eqref{tr1}.}\label{HPMRemoved}
	\end{figure}
	
	We consider the case representing the pricing of non-dividend paid European vanilla put options in a short term maturity. We take the following parameter values: risk-free interest rate $r=5\%$, volatility $\sigma=0.324336$, maturity date $T=6/12$, and strike price $E=40$. 
	
	We plot the exact solution \eqref{exact} and the approximations HPM1 and HPM2 at time $t=0$ in Fig.\ \ref{HPMRemoved}. We note that the first approximate solution HPM1 shown in dash-dotted line is indeed not smooth at one particular point, i.e., when the stock price $S$ is about the strike price $E$. This is different from the function HPM2, plotted as dashed line, that is smooth in its entire domain. Comparing them to the exact solution \eqref{exact}, we conclude that HPM2 is a better approximation and is in very good agreement with the exact solution.
	
	\begin{figure}[tbhp!]
		\centering
		\includegraphics[scale=0.4]{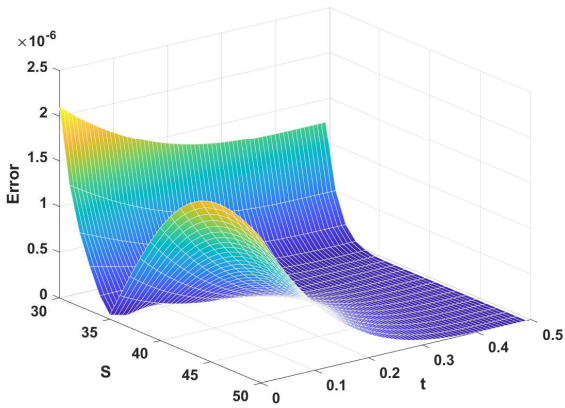}
		\\
		\caption{The error made by HPM2 in approximating the exact solution \eqref{exact}  of the single-asset put option when $T=6/12$.}\label{fig:Error}
	\end{figure}
	
	We also plot the difference between the price dynamics of the put options from the exact solution \eqref{exact} and the approximation HPM2 in Fig.\ \ref{fig:Error} with respect to the stock price $S$ and short term time to maturity date $t$. One can appreciate the accuracy of the pricing obtained using the homotopy perturbation method with the variable transformation we performed in this work.
	
	\subsection{Multi-asset basket options}\label{5b}
	
\begin{figure}[tbhp!]
		\mbox{}\hfill
		\includegraphics[scale=0.5]{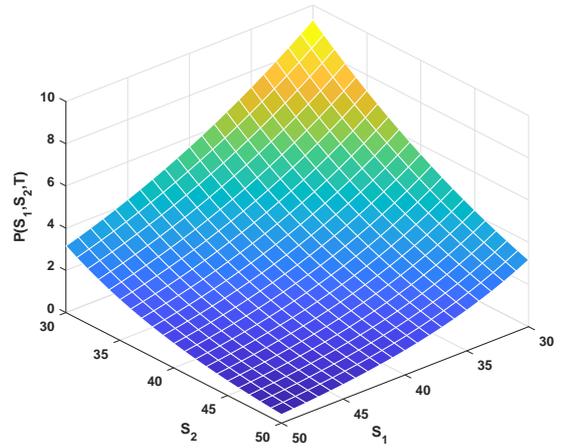}
		\hfill~
		\caption{{The basket put option value using $\sigma_1=0.1,\,\sigma_2=0.3,\,r=0.05,\, \text{and}\,\alpha_1=\alpha_2=0.5$.}}\label{fig:basketput}
	\end{figure}

	Next, we consider the multi-asset basket options with the exact solution given in Eq.\ \eqref{sol2}. The value is depicted in Fig.\ \ref{fig:basketput} as a function of the  first and second asset $S_1$ and $S_2$. 
	
\begin{figure}[tbhp!]
		\centering
		\includegraphics[scale=0.35]{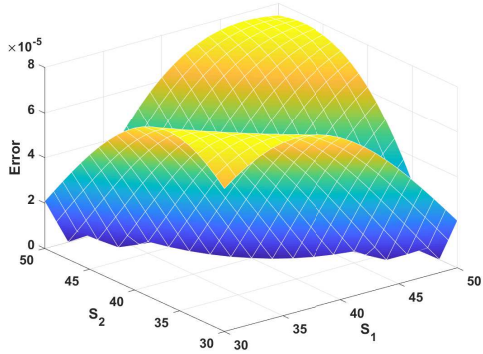}
		\\
		\caption{The error made by the homotopy perturbation method in approximating the exact solution \eqref{exact} when $T=6/12$.}\label{fig:Error2Asset}
	\end{figure}

	The error made by our approximate solution \eqref{tam4} using the homotopy method in approximating the analytical solution is depicted in Fig.\ \ref{fig:Error2Asset}, where it is clear that the series can provide a valuation of basket options rather accurately. 
	
	\subsection{Multi-asset quanto options}\label{sec5c}
	
\begin{figure}[tbhp!]
		\mbox{}\hfill
		\includegraphics[scale=0.5]{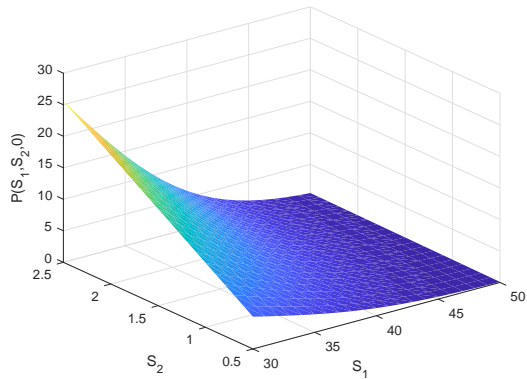}
		\hfill~
		\caption{The value of a quanto put option 
			for $\rho=1,\,\sigma_1=0.1,\,\sigma_2=0.3,r_1=0.03,\,\text{and}\,r_2=0.05$.}\label{fig:HPMQuanto}
	\end{figure}

	We plot the valuation of quanto options given by Eq.\ (\ref{eq:analyticquanto}) in Fig.\ \ref{fig:HPMQuanto}. The option 
	value decreases when the value of stock price $S_1$ increases for a fixed value of $S_2$. However, for a fixed option 
	stock price $S_1$, the value of options increases when the ratio of the exchange rate $S_2$ increases.

\begin{figure}[tbhp!]
		\mbox{}\hfill
		\includegraphics[scale=0.5]{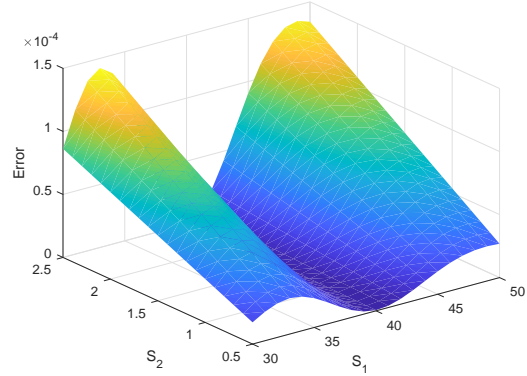}
		\hfill~
		\caption{The difference of a quanto put option between the analytical solution and the homotopy perturbation method using the same parameters as in Fig.\ \ref{fig:HPMQuanto}.}\label{fig:errorquanto}
	\end{figure}

	Using the homotopy perturbation method, our approximate solution is given in Eq.\ (\ref{eq:hpmquanto1}). The error made by our approximation compared to the exact solution is depicted in Fig.\ \ref{fig:errorquanto}. In general, the homotopy perturbation method can provide the valuation with high accuracy.
	
	\section{Conclusion}
	\label{sec6}
	
	We have considered the Black-Scholes equation for pricing put options and presented approximate solutions that were calculated using the homotopy perturbation method. In particular, we consider single-asset and multi-asset basket and quanto options. We showed that standardly applying the perturbation method will give a solution that is non-smooth (i.e., non-differentiable) at the strike price. However, by applying a variable transformation in advance, the method could be used to obtain a smooth approximation with a high accuracy. We have demonstrated excellent agreement of the approximation with the actual solution. Additionally, we also presented a novel transformation that changes a quanto multi-asset Black-Scholes equation into a single-asset one that allows us to obtain a solution that has not been reported before. 
	
	For future work, we will study the radius of convergence of our approximation. The applicability of the proposed transformation to solve, e.g., fractional or nonlinear Black-Scholes equations and compare them with numerical solutions is also proposed to be considered in the future.
	
	\section{Acknowledgement}
	The authors gratefully acknowledge Institut Teknologi Sepuluh Nopember, Surabaya, Indonesia, for financial support through the Visiting World Class Professor Programme year 2019. The authors are also grateful to the two anonymous reviewers and the editor for their feedbacks that greatly improved the manuscript. 
	
	\bibliographystyle{elsarticle-num}
	\bibliography{mybibfile}
	
\end{document}